\documentclass[twocolumn,floatfix,prl,showpacs]{revtex4}
\usepackage{graphicx}
\usepackage{amssymb}
\usepackage{amsmath}
\usepackage{color}
\usepackage{psfrag}
\usepackage{epsfig}
\usepackage{bbm}
\usepackage{bm}
\newcommand{\ket}[1]{| #1 \rangle}

\newcommand{\rb}[1]{\left( #1 \right)}

\newcommand{\ew}[1]{\langle #1 \rangle}
\newcommand{\beq}{\begin{eqnarray}}
\newcommand{\eeq}{\end{eqnarray}}

\newcommand{\op}[2]{| #1 \rangle \langle #2 |}

\newcommand{\eq}[1]{Eq.~(\ref{#1})}

\begin{document}
\title{Dark states in the magnetotransport through triple quantum dots}
\author{Clive Emary}
\affiliation{
  Institut f\"ur Theoretische Physik,
  Hardenbergstr. 36,
  TU Berlin,
  D-10623 Berlin,
  Germany
}

\date{\today}
\begin{abstract}
We consider the transport through a system of three coupled quantum dots in a perpendicular magnetic field. At zero field, destructive interference can trap an electron in a dark state --- a coherent superposition of dot states that completely blocks current flow.  The magnetic field can disrupt this interference giving rise to oscillations in the current and its higher-order statistics as the field is increased.  These oscillations have a period of either the flux-quantum or half the flux-quantum,  depending on the dot geometry.  We give results for the stationary current and for the shotnoise and skewness at zero and finite frequency.
\end{abstract}
\maketitle

The quantum-mechanical interference of electronic paths in a conductor gives rise to a number of interesting phenomena in mesoscopic physics.  Perhaps the most familiar is the occurrence of Aharonov-Bohm (AB) oscillations\cite{AB59} in the current through multiply-connected structures in a magnetic field \cite{aro87}.  These oscillations arise due to the accumulation of a phase difference $\phi$ between different paths through the device given by $\phi=\oint \mathbf{A}\cdot d\mathbf{l} = 2 \pi \Phi/\Phi_0$ with $\Phi$ the flux enclosed by the device and $\Phi_0=h/e$ the magnetic flux quantum.

The period of such oscillations is dependent on the nature of the interfering paths, and therefore on the specific system in question.  Flux periods of $\Phi_0$ are what one expects in conventional AB experiments, such as those on coherent beams of electrons in free space \cite{cha60}.  This period is also frequently encountered in mesoscopic experiments, for example, in normal metal rings \cite{web85} and in electronic Mach-Zehnder-style interferometers \cite{yac94,ji03}, including those with one \cite{yac95,sch97} or two \cite{hol01} quantum dots in the arms. 
The flux-period $\Phi_0/2$ is also observed, not just in superconducting systems \cite{par64}, but also in normal metals \cite{sha81,cha85} due to  weak-localisation effects \cite{aas81}.

A different quantum coherent effect was described for mesoscopic systems in Ref. \cite{mic06} --- that of coherent population trapping (CPT) in quantum dots.  In this all-electronic analogue of a quantum optics effect \cite{dark1,dark2,dark3}, the coupling geometry of a triple quantum dot (QD) leads to the establishment of a so-called ``dark-state'' that completely blocks the current through the device.  The dark-state is composed of a coherent superposition of electronic states in different dots.

\begin{figure}[t]
  \begin{center}
  \psfrag{1}{1}
  \psfrag{2}{2}
  \psfrag{3}{3}
  \psfrag{G1}{$\Gamma_1$}
  \psfrag{G2}{$\Gamma_2$}
  \psfrag{G3}{$\Gamma_3$}
  \psfrag{source}{source}
  \psfrag{drain}{drain}
  \psfrag{Phi}{$\Phi$}
  \psfrag{t12}{$t_{12}$}
  \psfrag{t13}{$t_{13}$}
  \psfrag{t23}{$t_{23}$}
  \epsfig{file=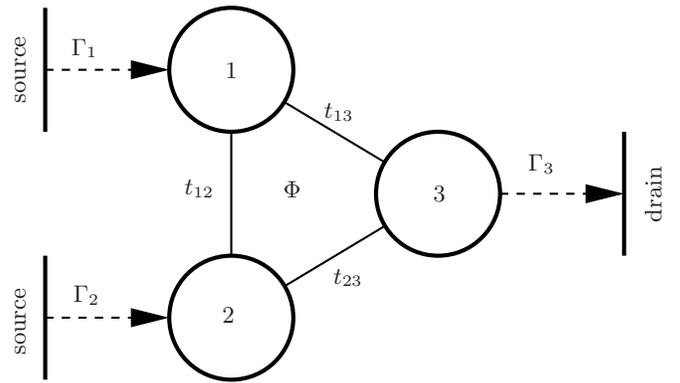, clip=true,width=1\linewidth}
  \caption{
    Three quantum dots are coupled coherently to one another via tunnel couplings $t_{ij}$, and incoherently to source and drain leads with rates $\Gamma_i$. Each dot contains a single level and by adjusting the relative positions of these levels, the system can be prepared in a dark state where no current flows despite the applied bias.
    In a perpendicular magnetic field, the structure encloses a magnetic flux $\Phi$ which causes a phase-difference between different paths around the system that can disrupt the dark state and lead to current flow.
    \label{schema}
   }
  \end{center}
\end{figure}

In this paper we consider the interplay of coherent population trapping and the AB phase.  We demonstrate how a magnetic field can destroy the delicate phase-cancelling that maintains the dark-state, lift the current blockade and give rise to current oscillations as the field increases.  Furthermore, we show that in a triple-QD structure such as in Fig.\ref{schema}, the oscillations can exhibit periods of both $\Phi_0$ and $\frac{1}{2}\Phi_0$ depending on the symmetry of the system.
We give results not only for the stationary current but also for the shotnoise and skewness (second and third current cumulants, respectively) both at zero \cite{gro06} and at finite frequency \cite{ce07}.
We also consider the effect of dephasing on the current oscillations.

We consider the QD geometry depicted in Fig. \ref{schema}. 
The AB effect requires that all three dots be coupled in a ring structure as shown, which is in contrast to Ref. \cite{mic06} where only two `bonds' were present, and no such effect would be observed.
We work in the strong Coulomb blockade regime such that there is at most one excess electron in the three dot system at any one time. Each dot has a single level relevant to transport, and we denote as $\ket{i}$ the state with an electron in dot $i$ .  The Hamiltonian in the basis $\left\{\ket{1},\ket{2},\ket{3}\right\}$ is then
\beq
  {\cal H} =
  \rb{
    \begin{array}{ccc}
      \Delta & t_{12}e^{i\phi} & t_{13} \\
      t_{12}e^{-i\phi} & -\Delta & t_{23} \\
      t_{13} & t_{23} & \epsilon
    \end{array}
  }
  ,
  \label{H}
\eeq
where $\epsilon$ and $\Delta$ describe the energies of the dot levels and $t_{ij}$ are the tunnel couplings.  Without magnetic field, time-reversal symmetry means that all $t_{ij}$ are real, and we take them all to be positive.  Application of the magnetic field breaks this symmetry and the amplitudes will be complex in general.  We choose a gauge such that the phase $\phi= 2 \pi \Phi/\Phi_0$ is accumulated on the bond between dots 1 and 2.  Finally, dots 1 and 2 are connected to source leads, and dot 3 to the drain.

The density matrix (DM) for the system $\rho(t)$ contains entries not only for the three single electron states $\ket{i}$, but also the empty state $\ket{0}$. Within the Born-Markov and infinite-bias approximations, the time-evolution of the DM is given by the generalised master equation in the Lindblad form,
\beq
  \frac{d\rho}{dt}=-i[{\cal H},\rho]
  +\sum_{k}
    D_{k}^{\vphantom{\dagger}}\rho D_{k}^{\dagger}
  -\textstyle{\frac{1}{2}}D_{k}^{\dagger}D_{k}^{\vphantom{\dagger}}\rho
  -\textstyle{\frac{1}{2}}
    \rho D_{k}^{\dagger}D_{k}^{\vphantom{\dagger}}
    ,
  \label{master}
\eeq
where the quantum jump operators 
$D_1 = \sqrt{\Gamma_1} \op{1}{0}$,
$D_2 = \sqrt{\Gamma_2} \op{2}{0}$, and
$D_3 = \sqrt{\Gamma_3} \op{0}{3}$ describe irreversible tunnelling of electrons into and out of the system with rates $\Gamma_i$.  In the following we set all these rates to equal: $\Gamma_i=\Gamma$.

Starting from initial DM $\rho(0)=\op{0}{0}$, the DM at subsequent time has ten non-zero elements if we assume the most general parameters in ${\cal H}$.  These elements we arrange into the column vector 
$\bm{\rho}(t) = \rb{
  \rho_{00},\rho_{11},\rho_{22},\rho_{33},
  \Im\rho_{12},\Im\rho_{13},\Im\rho_{23},
  \Re\rho_{12},\Re\rho_{13},\Re\rho_{23}
}^T$.  
The master equation 
(\ref{master}) can then be written 
\beq
  \dot{\bm{\rho}} = M \bm{\rho}
  ,
\eeq
with the time-evolution matrix $M$ given in the Appendix.
The stationary properties of the system are determined by the eigenvalues and eigenvectors of $M$\cite{ce07}.  The stationary DM, $\rho(\infty)$ is given by the eigenvector of $M$ with zero eigenvalue, whence the stationary average current $\ew{I} = \Gamma \rho_{33}(\infty)$.  To calculate higher-order statistics we require the full spectral decomposition of $M$ \cite{ce07}.
We initially consider the behaviour of the system in the absence of dephasing and return its effects subsequently.

\section{zero-field current}

We first consider the properties of the system at zero magnetic field ($\phi=0$).  In Ref.~\cite{mic06}, it was shown that, with the special choice of parameters
$t_{12}=0$, $\epsilon=\Delta=0$ and $t_{13}=t_{23}$, the system always reaches the trapped state $\ket{\Psi}=\rb{\ket{1}-\ket{2}}/\sqrt{2}$ in the stationary limit (in the absence of dephasing).  This exact parameter set is unlikely to pertain in experiment, and it is important to show the existence of the dark state for more general parameters.

If we set  $\Delta=\Delta_0$ with
\beq
  \Delta_0 \equiv \frac{t_{12}}{2 t_{13}t_{23}}\rb{t_{13}^2-t_{23}^2}
\eeq
then, as can easily be verified, the state
\beq
  \ket{\Psi_\mathrm{dark}} = 
  \frac{1}{\sqrt{t_{13}^2 + t_{23}^2}}
  \rb{t_{23} \ket{1}- t_{13} \ket{2}}
\eeq
is an eigenstate of Hamiltonian (\ref{H}).  Moreover, the vector corresponding to the pure DM
\beq
  \rho_\mathrm{dark} = \op{\Psi_\mathrm{dark}}{\Psi_\mathrm{dark}}
\eeq
is the eigenvector of $M$ with eigenvalue zero, and thus $\rho_\mathrm{dark}$ is the stationary state of the system.  Since this state has no electronic density at the drain dot (dot~3), the stationary current through the device is exactly zero.
If we assume therefore, that we have experimental control of the detuning $\Delta$ by, for example, backgates under the dots, then a dark-state always can always be found at zero field by sweeping $\Delta$.  
It should be noted that the detuning $\epsilon$ does not affect the existence of the dark-state.

\begin{figure}[t]
  \psfrag{DD}{$(\Delta-\Delta_0)/\Gamma$}
  \psfrag{IG}{$\ew{I}/\Gamma$}
  \begin{center}
  \epsfig{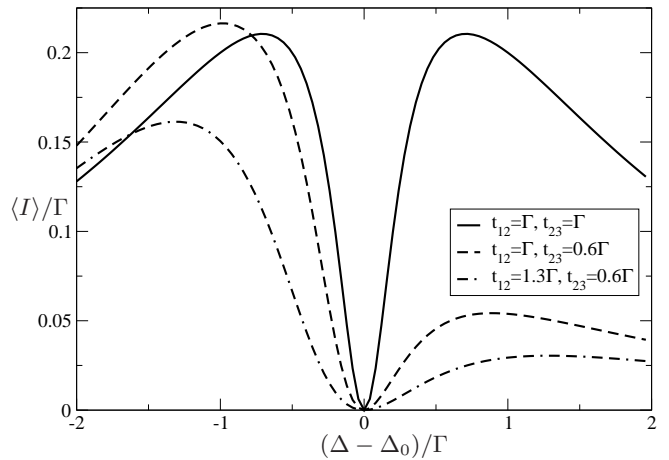}
  \caption{
    At zero-field,
    the stationary current $\ew{I}$ through the three-dot system shows a pronounced anti-resonance with complete current blocking at a detuning of $\Delta=\Delta_0$ where the dark-state forms.  The three curves show the current for different values of the coupling parameters $(t_{12},t_{23})$, with $t_{13}=\Gamma$ in each case.  At zero-field, a dark state always exists whenever both $t_{13}$ and $t_{23}$ are finite.
    \label{B0}
   }
  \end{center}
\end{figure}

A special instance of our geometry is when $t_{13}=t_{23} =t_0$.  In this case, the dark-state forms at a value of the detuning $\Delta=\Delta_0=0$. With this choice, the system is symmetric under the exchange of dots 1 and 2, and correspondingly we will refer to this situation as the `symmetric' case in what follows.  This special case admits a number of exact solutions.
 
In this symmetric case at zero field with $\epsilon=0$ and $\Delta$ a freely variable parameter, we have for the stationary current 
\beq
  \ew{I}= \frac{4 \Gamma \Delta^2 t_0^2}
  {4 \Delta^4 + \Gamma^2 t_{12}^2 + 4(t_0^2-t_{12}^2)^2 
  + \Delta^2(\Gamma^2 + 6 t_0^2 + 8t_{12}^2)}
.
\nonumber
\eeq
Figure \ref{B0} shows the stationary current through the device for $B=0$ as a function of the detuning $\Delta$. We show not only this result for the symmetric case, but also numerical results for the current with various different couplings.  In each case an anti-resonance occurs with complete current suppression at $\Delta = \Delta_0$.
In the following we will always set $\Delta = \Delta_0$, unless otherwise stated, such that there is CPT at zero-field.

\section{Current oscillations}

\begin{figure}[t]
  \psfrag{pp}{$\Phi/\Phi_0$}
  \psfrag{II}{$\bar{I}$}
  \begin{center}
  \epsfig{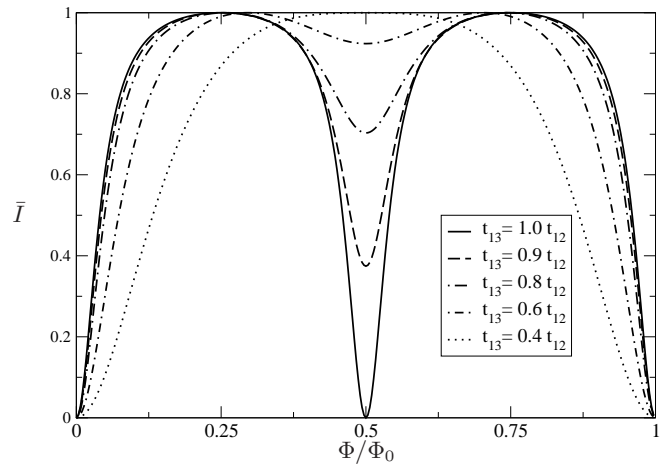}
  \caption{
    The stationary current through the three-dot system shows pronounced oscillations as a function of the applied flux.  
    Plotted here is $\bar{I}=\ew{I}/\ew{I}_\mathrm{max}$, the ratio of the current to its maximum value, which always occurs at $\Phi=\frac{1}{4}\Phi_0$.
    The different curves are for different values of $t_{13}$. Other parameters were $t_{12}=t_{23} = \Gamma$, $\epsilon=0$ and $\Delta=\Delta_0$, such that CPT occurs at zero field.
    In the symmetric case with $t_{13}=t_{23}$ (solid curve), CPT trapping occurs at $\frac{n}{2}\Phi_0$, $n=0,1,2,\ldots$ and the flux-period of the oscillations is thus $\frac{1}{2}\Phi_0$.  As $t_{13}$ moves away from symmetry, the dark-state current blocking at $\frac{n}{2}\Phi_0$, $n$ odd disappears, and the period of the oscillations doubles to $\Phi_0$.
    \label{figt13}
   }
  \end{center}
\end{figure}

The application of magnetic field has the capacity to lift the dark-state current blockade and give rise to oscillations in the current.
In the symmetric case, an exact expression for the current at finite field can be found.  With $\epsilon=0$, we have
\beq
  \ew{I}\rb{\phi}= \frac{4\Gamma t_0^2 t_{12}^2 \sin^2\phi }
  {4 (t_{12}^4+t_0^4) +t_{12}^2(\Gamma^2-t_0^2(1+7\cos2\phi))}
  ,
\eeq
which is plotted in Fig.~\ref{figt13}.  The current shows clear oscillations with superconducting flux period $\frac{1}{2}\Phi_0$.  We will return to a discussion of which paths interfere here in a moment. Let us first note that the maximum current as a function of $\Phi$ occurs at $\Phi = \frac{1}{4}\Phi_0$, irrespective of $t_{12}$, and that the current at this value of $\Phi$ is itself maximised by setting $t_{12}=t_0$.  With this choice, the current in the weak coupling limit $t_0 \ll \Gamma$ is given by $\ew{I}=4 t_0^2/\Gamma \sin^2 \phi$, and in the opposite regime, $t_0 \gg \Gamma$, we have
$\ew{I}=2\Gamma/7$ for all $\phi$ except at $n \pi$ where it is exactly zero.  This means that the oscillations are easier to observe in the weak coupling limit.

The suppression of the current at $\Phi = \pm \frac{n}{2}\Phi_0$, with $n$ odd, is a consequence of the additional symmetry of the above situation. 
Figure \ref{figt13} also shows the behaviour of the current as we move away from the symmetric coupling.  As the asymmetry increases, the features at odd multiples of $\frac{1}{2}\Phi_0 $ disappear, doubling the period of the oscillations to $\Phi_0$.
Setting $t_{12}=t_{23}$ and $t_{13}=t_{23}\rb{1+\alpha}$, a perturbation series for the current in $\alpha$ away from the symmetric case, shows the current at $\Phi=\frac{1}{2}\Phi_0$ to be
\beq
  \ew{I} = 16 t_{23}^2 \alpha^2 /\Gamma + O(\alpha^4)
  .
\eeq
That this dependence is quadratic suggests that some degree of current suppression at odd-multiples of $\frac{1}{2}\Phi_0$ may be visible in experiment.

The above results can be understood by consideration of the interference between different paths around the dots.  Consider the three dot system to be occupied and isolated from the leads.  If we assume that at time $\tau=0$ the system is in the pure state $\ket{\Psi(0)}$, then the wavefunction at later time $\tau$ is  
$\ket{\Psi(t)} = e^{-i{\cal H}t} \ket{\Psi(0)}$ which, for small times,
can be expanded as $\ket{\Psi(t)}\approx \rb{1 -i \tau {\cal H} - \tau^2 {\cal H}^2} \ket{\Psi(0)}$.
Consider the system initially in the state $\ket{1}$.  To first order is $\tau$, evolution under the full Hamiltonian with $\Delta=\Delta_0$ gives ${\cal H}\ket{1} = \Delta_0 \ket{1} + t_{12} e^{-i \phi} \ket{2} + t_{13}\ket{3}$.  Therefore the first-order amplitude for the transmission from dot 1 to 3 is $a^{(1)}_{31} = \ew{3|{\cal H}|1}=t_{13}$.  Similarly, the amplitude from dot 2 to 3 is $a^{(1)}_{32} = t_{23}$.  Thus, if we start the system in the dark-state superposition $1/N\rb{t_{23}\ket{1}-t_{13}\ket{2}}$, with norm $N=\sqrt{t_{23}^2 + t_{13}^2}$, these two paths interfere destructively at dot 3 with total amplitude $a^{(1)} = (t_{23} a^{(1)}_{31} - t_{13}a^{(1)}_{32})/N=0$.  The dark state is therefore stabilised against first-order tunnelling regardless of applied field.

The unblocking of the system at finite $B$-field occurs at second-order.  Consider the second-order amplitude from 1 to 3:
$a_{31}^{(2)} = (\Delta_0 t_{13} + \epsilon t_{13} + e^{-i \phi} t_{12}t_{23})/N^2$ which has contributions from the three paths 113, 133, and 123.  Similarly, the three paths 223, 233 and 213  give the amplitude from dot 2 to 3 as
$a_{32}^{(2)} = (-\Delta_0 t_{23} + \epsilon t_{23} + e^{i \phi} t_{12}t_{13})/N^2$.  The total second-order amplitude $a^{(2)}$ for the dark-state electron to tunnel to dot 3 is
$
 a^{(2)} = t_{12} 
 \rb{e^{-i\phi}-1}
 \rb{e^{i\phi} t_{13}^2 + t_{23}^2}/N^2
$.  
The corresponding probability $p^{(2)} = |a^{(2)}|^2$ is
\beq
 p^{(2)} = 2 t^2_{12} 
 \rb{1-\cos \phi}
 \rb{t_{13}^4 + t_{23}^4 + 2 t^2_{13} t^2_{23} \cos \phi}/N^4
 .
\eeq
We see  immediately that this probability is zero at zero field.  Furthermore, if $t_{13}^2 \ne t_{23}^2$, the probability $p^{(2)}$, and hence the current, has period $\Phi_0$.  However, if 
$t_{13}^2 = t_{23}^2$, this probability becomes $p^{(2)} =4 t^2_{12} t_{13}^4 \sin^2\phi$, which gives rise to current blocking with period $\frac{1}{2}\Phi_0$.
This halved oscillation period results from the symmetry of the two paths 123 and 213 when $t_{13}=t_{23}$.

This symmetry is reflected in the detuning required to obtain the dark state.  When $t_{13}=t_{23}$, we have $\Delta_0=0=-\Delta_0$.  However, when we have a coupling asymmetry, $\Delta_0\ne 0\ne-\Delta_0$, and the period is doubled.  Then setting $\Delta=-\Delta_0$, instead of $\Delta=\Delta_0$, is equivalent to shifting the phase of the current oscillations by $\phi$, such that the dark-states occur at $\frac{n}{2}\Phi_0$; $n$ odd.  On the other hand, creation of a dark state at other values of $\Phi$ is impossible since this would require complex values of $\Delta$, which is obviously a real parameter.

\section{dephasing}
We model the influence of dephasing due to charge noise through the introduction of the three jump operators
\beq
  D^\gamma_i = \sqrt{\gamma} \op{i}{i};\quad i=1,2,3
  ,
\eeq
which enter in \eq{master} in the same way as do the jump operators $D_i$ above.  Out of simplicity, we assume the decoherence rate $\gamma$ to be the same for each dot and we give here only results for the case when all three couplings are equal, $t_{ij}=t$.  In this case, the current as a function of the phase $\phi$ and decoherence rate $\gamma$ is
\beq
  \ew{I}(\phi,\gamma) = 
  \frac{
    \Gamma
    (
      4 \gamma (\Gamma+2 \gamma) t_0^2
      + 2 f(\phi)
    )
  }
  {
    \Gamma\gamma(\Gamma+2\gamma)^2
    + 2 (2 \Gamma^2 + 13 \Gamma \gamma + 14 \gamma^2) t_0^2
    + 7 f(\phi)
  },
  \nonumber
\eeq
with $\phi$-dependence contained in the function
\beq
  f(\phi) = 8 t_0^4 (\Gamma + 3 \gamma) (\Gamma+2\gamma)^{-1}\sin^2 \phi
  .
\eeq
Since $f(0)=0$ ,the dephasing leads to a finite current at zero-field through the disruption of the coherence between the two dark-state dots \cite{mic06}.  The effect of dephasing on the oscillations can be quantified through the visibility:
\beq
  \nu(\gamma) \equiv
  1 - 
  \frac{
  \ew{I}(\phi_\mathrm{min},\gamma)
  }
  {
  \ew{I}(\phi_\mathrm{max},\gamma)
  }
  .
\eeq
This is found to be
\beq
  \nu(\gamma) =
  \frac{4 \Gamma (\Gamma + 3 \gamma)t_0^2}
  {\Gamma\gamma(\Gamma+2 \gamma)^2 
    + 2 t_0^2 (2 \Gamma^2 + 13 \Gamma \gamma + 14 \gamma^2)
  }
  .
\eeq
For small dephasing, $\gamma \ll \Gamma$, the visibility deviates from unity as
 \beq
   \nu = 1 - \gamma
   \rb{
     \frac{7}{2\Gamma}
     +\frac{\Gamma}{4 t_0^2}
   }
 \eeq
and for strong dephasing $\gamma \gg \Gamma$, we have $\nu = 3 t_0^2 / \gamma^2$.

\section{Higher-order current statistics}

\begin{figure}[t]
  \begin{center}
  \psfrag{PP}{$\Phi/\Phi_0$}
  \psfrag{F0}{$F^{(2)}$}
  \epsfig{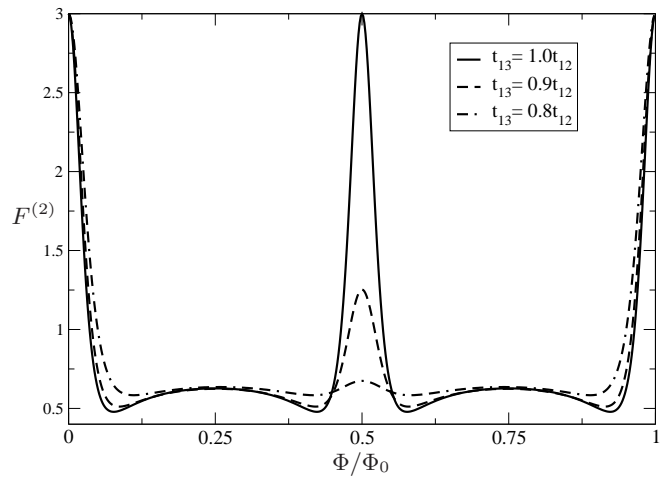}
  \caption{  
    The zero-frequency Fano factor $F^{(2)}(0)$ for the three dot system as a function of magnetic flux also exhibits oscillations.  We observe strong superPoissonian peaks with $F^{(2)}(0)=3$ at the values of flux for which the dark-state forms \cite{gro06}.  Away from these points, the noise is subPoissonian.  The period of these oscillations is the same as for the current.  Same parameters as for Fig.~\ref{figt13} except for the displayed values of $t_{13}$.
    \label{F2fig}
   }
  \end{center}
\end{figure}

The dark state and magnetic field also leave their mark on the higher statistics of the current through the device.  We consider here the two finite-frequency Fano-factors: $F^{(2)}(\omega) \equiv S^{(2)}(\omega) /\ew{I}$ where  $S^{(2)}(\omega)$ is the shotnoise, and $F^{(3)}(\omega,\omega') \equiv S^{(3)}(\omega,\omega') /\ew{I}$ with the skewness
\beq
  S^{(3)}(\omega,\omega') = \int d \tau  d \tau' 
  e^{i \omega \tau +i \omega' \tau' }
  \ew{\delta I (0) \delta I (\tau)\delta I (\tau')}
  .
\eeq
These quantities can be straightforwardly calculated\cite{ce07} from $\lambda_i$, the eigenvalues of $M$, and $V$ the corresponding matrix of its eigenvectors.   Let ${\cal L}_J$ be the jump operator that transfers an electron to the drain from dot 3.  In the basis of the vector $\mathbf{\rho}$, it has elements $({\cal L}_J)_{ij}= \Gamma \delta_{i1} \delta_{j4}$.  The Fano factors can then be expressed solely in terms of the  eigenvectors $\lambda_i$ and the quantities
$
  c_k \equiv (V^{-1}{\cal L}_J V)_{kk}
$.

The finite-frequency shotnoise Fano factor is given by
\beq
  F^{(2)}(\omega) 
  &=& 1 - 2\sum_k\frac{c_k
  \lambda_k}{\omega^2+\lambda_k^2}.
 \label{F2}
\eeq
The zero-frequency result, $F^{(2)}(0)$, is shown in Fig.~\ref{F2fig} as a function of $\Phi$, for several couplings.  At $\Phi=n\Phi_0$, $n=0,1,\ldots$,  we see highly superPoissonian maxima with Fano factor $F^{(2)}(0)=3$, which is the same value as found for the $t_{12}=0$ model discussed in Ref.~\cite{gro06}.  In the symmetric case $t_{13}=t_{23}$, we see further maxima, at $\Phi = \frac{n}{2}\Phi_0$, also with $F^{(2)}(0)=3$.  These latter disappear as the coupling asymmetry increases.  In between these sharp superPoissonian peaks, the shotnoise is strongly subPoissonian, further illustrating the dramatic variation in systems behaviour as the field is changed.

\begin{figure}[t]
  \begin{center}
  \psfrag{pp}{$\Phi/\Phi_0$}
  \psfrag{w}{$\omega$}
  \epsfig{file=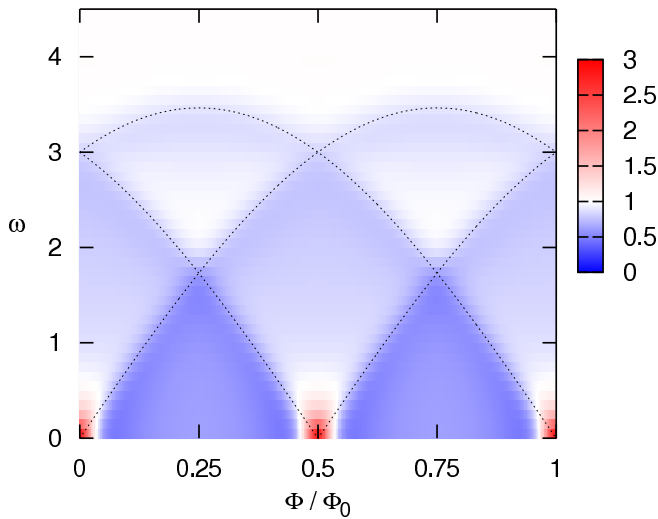, clip=true,width=1\linewidth}
  \caption{(colour online)
    Contour plot of the finite-frequency Fano factor $F^{(2)}(\omega)$ as a function of magnetic flux $\Phi$ and frequency $\omega$.  Colours white, red, and blue correspond to Poissonian, super-, and sub-Poissonian values respectively.  
    Large superPoissonian peaks occur in the zero frequency limit only.  The shotnoise also shows a series of inflexion points at a set of frequencies corresponding to the energy differences $\Delta E_{ij}$ of the Hamiltonian (dotted lines).
   Same parameters as Fig.~\ref{figt13} with all $t_{ij}=\Gamma$.
    \label{F2w}
   }
  \end{center}
\end{figure}

The shotnoise Fano factor at finite frequency $F^{(2)}(\omega)$ is shown for symmetric coupling in Fig.~\ref{F2w}. 
We see that the large superPoissonian peaks occur only close to zero-frequency, as the majority of the behaviour is subPoissonian.
Nevertheless, further structure is to be observed at finite frequency, with a number of inflexion points occurring as a function of $\omega$, the locations of which are determined by the spectrum of the isolated Hamiltonian ${\cal H}$.  The inflexions are located at $\omega=\Delta E_{ij}$, where $\Delta E_{ij} = |E_i-E_j|$ are the differences between all the eigenenergies of ${\cal H}$.  In the case where all three couplings are equal, these energies are obtained from the three solutions of the equation $E_i^3 - 3t_0^2 E_i - 2 t_0^3 \cos \phi=0$, and the corresponding differences are shown overlaid on Fig.~\ref{F2w}.

\begin{figure}[t]
  \begin{center}
  \psfrag{pp}{$\Phi/\Phi_0$}
  \psfrag{F3}{$F^{(3)}$}
  \epsfig{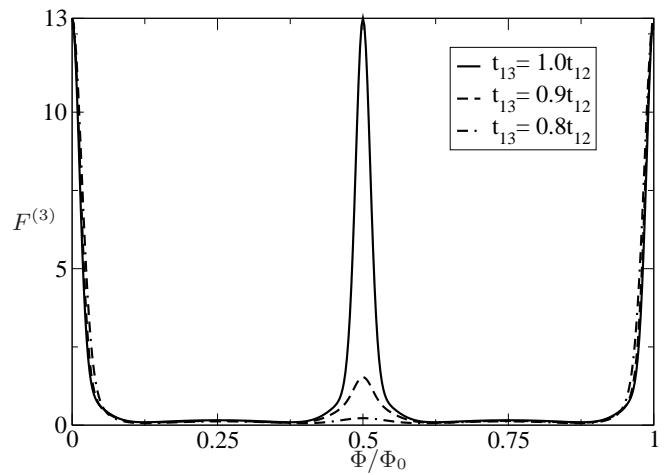}
  \caption{
     The zero-frequency skewness Fano-factor $F^{(3)}(0)$ as a function of flux $\Phi$.  The behaviour is similar to that of the shotnoise, with the highly superPoissonian maximum value of $F^{(3)}(0)_\mathrm{max}=13$. Parameters as in Fig.~\ref{F2fig}.
    \label{F30}
   }
  \end{center}
\end{figure}

The skewness is calculated from an expression similar to \eq{F2}, but lengthier.
We obtain the zero-frequency result shown in Fig.~\ref{F30}, which has the zero-field limit of $F^{(3)}(0) = 13$, which is again the same as in Ref.~\cite{gro06}.  The behaviour of the skewness as a function of field strength is similar to that of the shotnoise, but here the contrast between the values with and without the dark start is even more pronounced.

\begin{figure}[t]
  \begin{center}
  \psfrag{wd}{$\omega'0$}
  \psfrag{w}{$\omega$}
  \epsfig{file=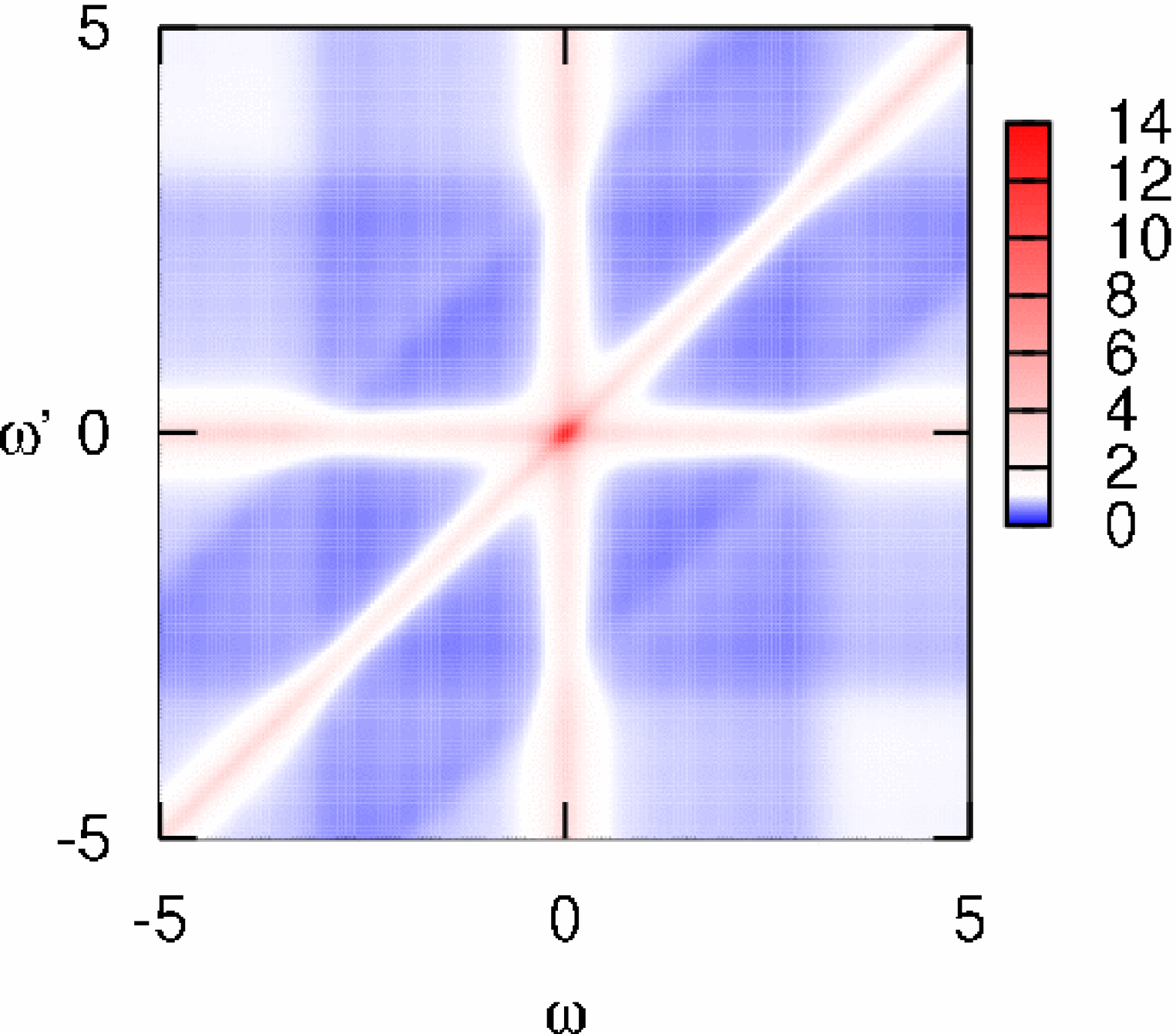, clip=true,width=0.48\linewidth}
  \epsfig{file=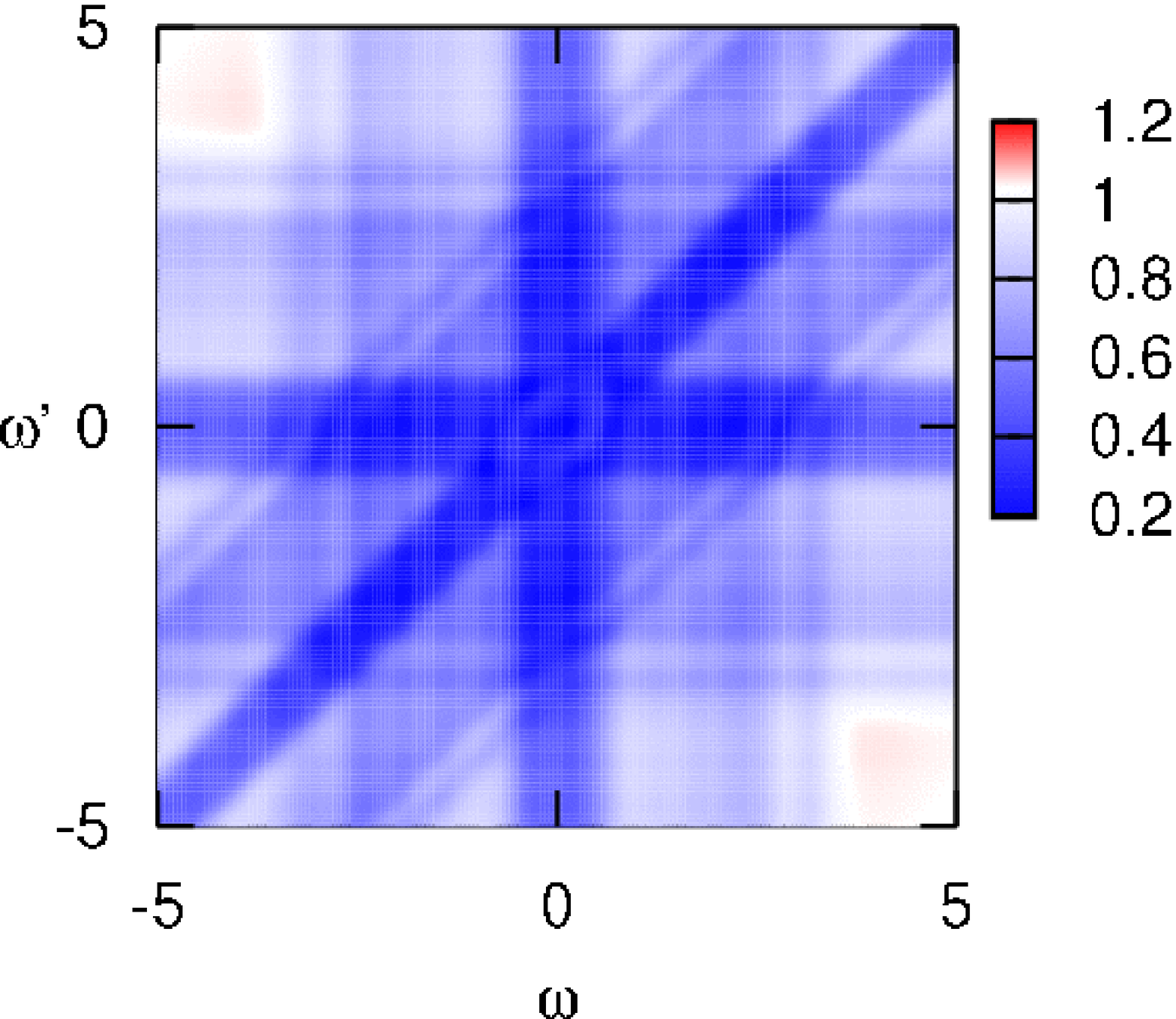, clip=true,width=0.48\linewidth}
  \epsfig{file=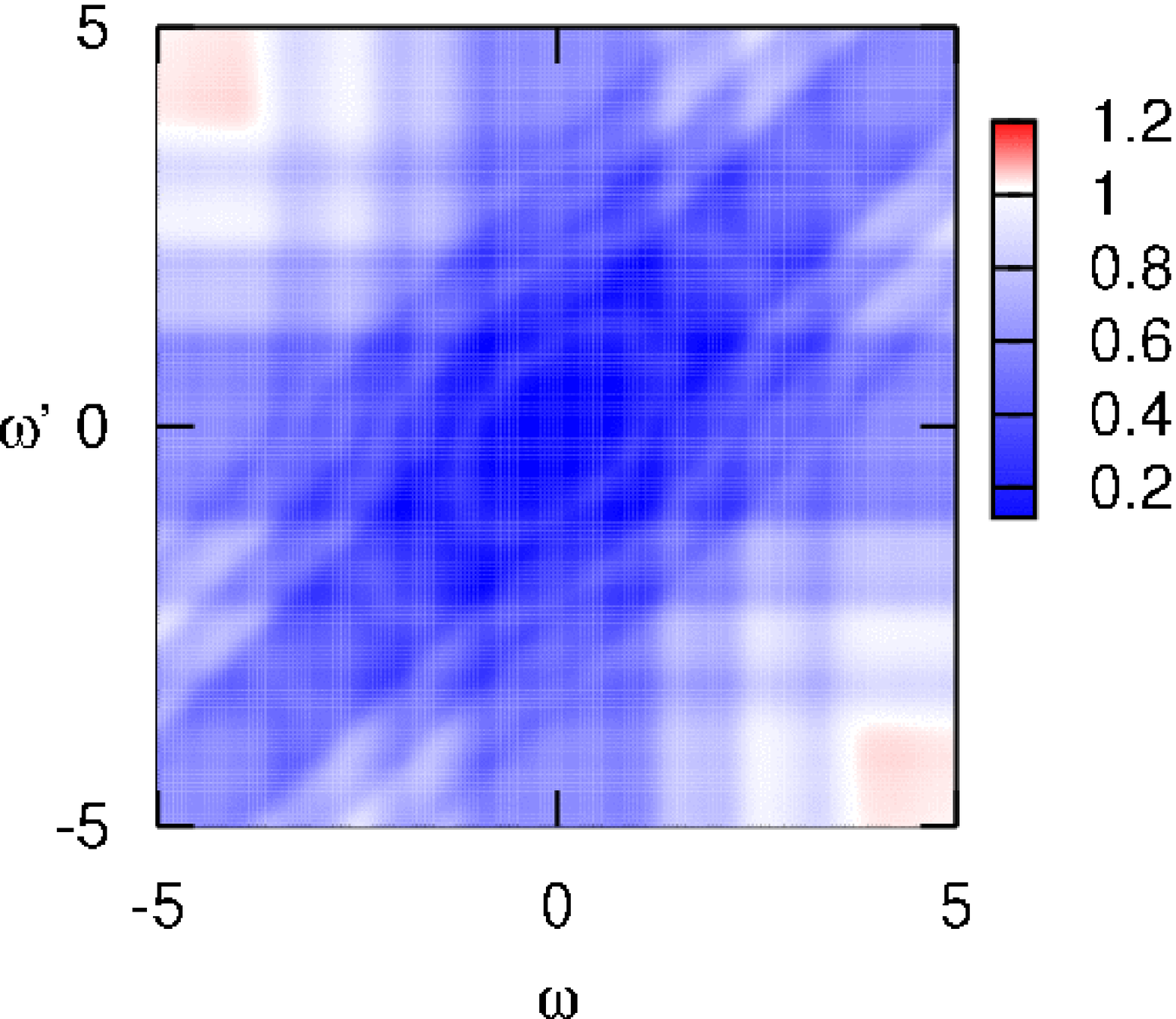, clip=true,width=0.48\linewidth}
  \epsfig{file=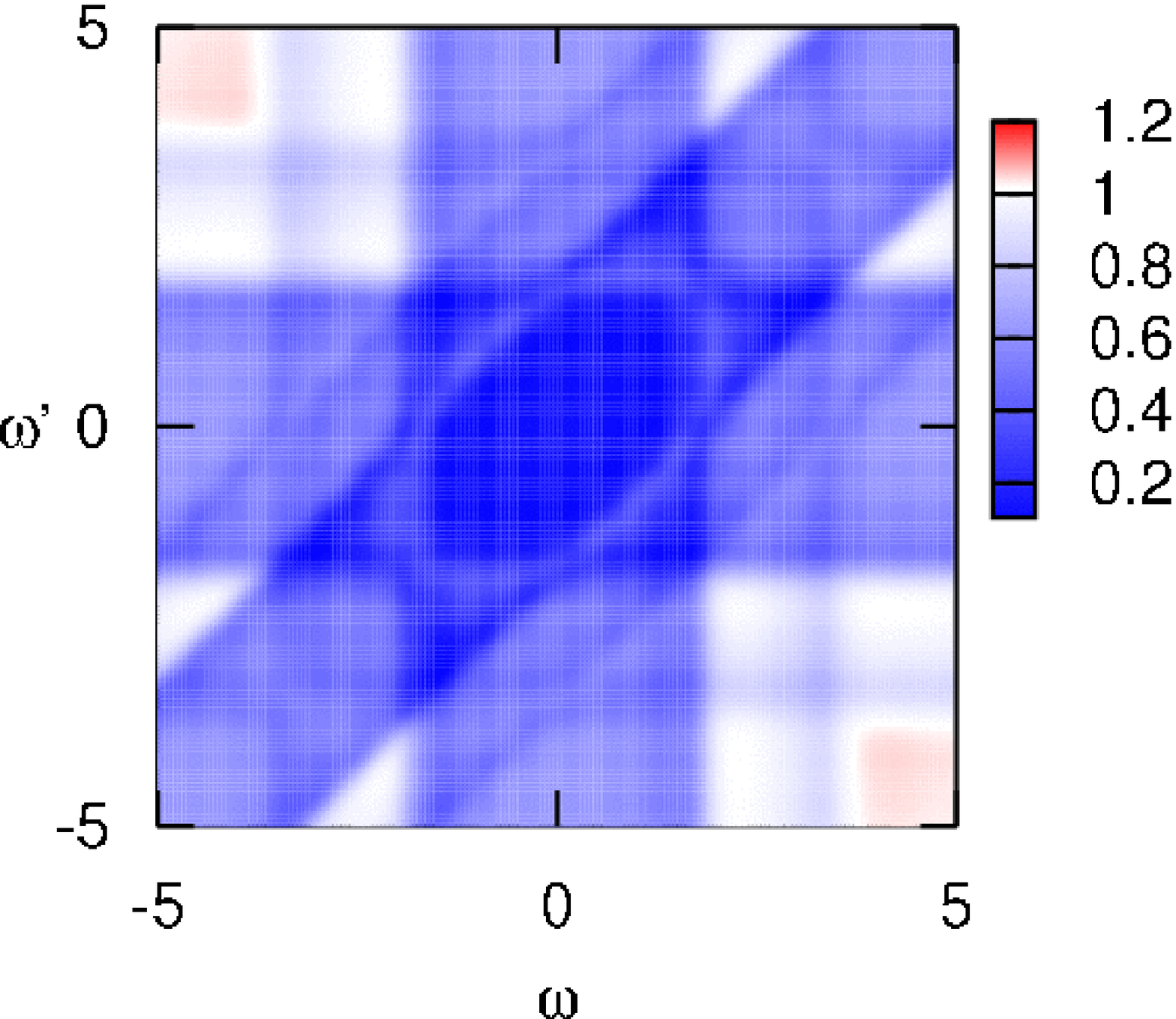, clip=true,width=0.48\linewidth}
  \caption{(colour online)
    Contour plots of the finite-frequency skewness Fano-factor $F^{(3)}(\omega)$ as a function of its two frequency arguments $\omega$ and $\omega'$ for values of the magnetic flux $\Phi/\Phi_0 =0,1/12,1/6,1/4$. Parameters and colour scheme as in Fig \ref{F2w}.  Strong superPoissonian behaviour only occurs close to $\Phi/\Phi_0 = \textstyle{\frac{n}{2}}$, otherwise the skewness is predominantly subPoissonian.  The fine structure arises from resonance between the frequencies  $|\omega|$, $|\omega'|$ and $|\omega-\omega'|$ and the level-splittings of the isolated dot-system.
    \label{F3w}
   }
  \end{center}
\end{figure}

In Fig.~\ref{F3w} we plot the finite-frequency skewness for several values of magnetic field with symmetric couplings.
For $\Phi=\textstyle{\frac{n}{2}}\Phi_0$; $n=0,\pm 1,\ldots$ in this symmetric case, the skewness shows a sharp superPoissonian peak at the origin and also strong superPoissonian behaviour along the symmetry lines of $F^{(3)}$.  Away from these values of the magnetic field, the skewness is strongly subPoissonian, except for slight SuperPoissonian shoulders at intermediate frequencies before the Poissonian value of unity is recovered at high frequency \cite{ce07}.  The fine structure in the skewness arises from resonances between the three frequencies $|\omega|$, $|\omega'|$ and $|\omega-\omega'|$ and the energy differences $\Delta E_{ij}$. 

\section{conclusion}
The behaviour of the coupled triple quantum dot system in a perpendicular magnetic field studied here is governed by the interplay of two quantum-coherent effects: coherent population trapping and the Aharonov-Bohm phase.

We have shown that a dark-state, for which no current flows, exists at zero field for arbitrary couplings between the dots provided $t_{13}$ and $t_{23}$ are both finite.
The magnetic field can lift destructive interference maintaining the dark state, and give rise to oscillations in the current.  For arbitrary parameters the period of these oscillations is $\Phi_0$, but in the special case when the coupling strengths $t_{13}$ and $t_{23}$ are equal the period is halved to $\frac{1}{2}\Phi_0$.
These oscillations are also visible in the zero-frequency shotnoise and skewness which show large oscillations between strong superPoissonian and subPoissonian behaviour.  Finally, at finite frequency these quantities show considerable structure which again show dramtic dependence of the magnetic field.

\section{acknowledgements}
This work was supported by the WE Heraeus foundation and DFG grant BR 1528/5-1.  I am grateful to T.~Brandes, R.~Haug, and M.~Rogge for useful discussions.

\appendix
\section{Appendix: Time-evolution matrix}

The equation-of-motion matrix for the system is
\begin{widetext}
\beq
  M = 
  \rb{
    \begin{array}{cccccccccc}
     -2 \Gamma & 0 & 0 & \Gamma & 0 & 0 & 0 & 0 & 0 & 0 \\
     \Gamma & 0 & 0 & 0 & -2t_{12}\cos\phi & -2t_{13} & 
         0 & 2t_{12}\sin\phi & 0 & 0 \\
     \Gamma & 0 & 0 & 0 & 2t_{12}\cos\phi & 0 & -2t_{23} & 
         -2t_{12}\sin\phi & 0 & 0 \\
     0 & 0 & 0 & -\Gamma & 0 & 2 t_{13} & 2 t_{23} & 0 & 0 & 0 \\
     0 & t_{12}\cos\phi & -t_{12}\cos\phi & 0 & -\gamma & 0 & 0 & 
         2\Delta &  t_{2 3} & -t_{13} \\
     0 &  t_{13}& 0 & -t_{13} & 0 & -\Gamma/2 -\gamma& t_{12}\sin\phi& 
          t_{2,3} &  -\Delta + \epsilon&  -t_{12}\cos\phi\\
     0 & 0 &  t_{23}& -t_{23} & 0 &  -t_{12}\sin\phi &  -\Gamma/2 -\gamma&  
          t_{13}& -t_{12}\cos\phi & -\Delta + \epsilon \\
     0 & -t_{12}\sin\phi & t_{12}\sin\phi & 0 & 2\Delta  & -t_{23} &  
          -t_{13}& -\gamma & 0 & 0 \\
     0 & 0 & 0 & 0 & -t_{23} &  \Delta - \epsilon& t_{12}\cos\phi & 
          0 & -\Gamma/2-\gamma &  t_{12}\sin\phi \\
     0 & 0 & 0 & 0 & t_{13} & t_{12}\cos\phi & \Delta - \epsilon & 
          0 &  -t_{12}\sin\phi & -\Gamma/2 -\gamma
    \end{array}
  }
  \label{M}
\eeq
\end{widetext}



\end{document}